\PassOptionsToPackage{hyphens}{url}

\documentclass[%
    format=sigconf,
    natbib=true,
    anonymous=false,
]{acmart}

\setcopyright{none}
\renewcommand\footnotetextcopyrightpermission[1]{}
\usepackage{url}
\usepackage{booktabs}
\usepackage{tabularx}
\usepackage{threeparttable}
\usepackage{mdframed}
\usepackage{multirow}
\usepackage{tikz}
\usepackage{rotating}
\usepackage{enumitem}
\usepackage{listings}
\usepackage{csquotes}

\setlist[description]{leftmargin=0pt}
\setlist[itemize]{leftmargin=12pt}

\usetikzlibrary{trees,shapes.geometric,shapes.symbols,chains,calc,positioning,arrows,decorations.pathmorphing,decorations.pathreplacing}

\newlength\bubblesize
\newcommand\yes{\tikz[baseline=0.1ex] \fill[black] (\bubblesize,\bubblesize) circle (\bubblesize);}

\newcommand\no{\tikz[baseline=0.1ex] \draw[black, line width=0.2ex] (\bubblesize,\bubblesize) circle (\bubblesize-.5\pgflinewidth);}

\newcommand{\studyinvites}{91,177}
\newcommand{\studyvalid}{70}
\newcommand{\surveyinvites}{62,462}
\newcommand{\surveyvalid}{308}
\newcommand{\allanalyzedapps}{1,762,868}
\newcommand{\obfuscatedapps}{439,232}
\newcommand{\toolname}{\textsc{Obfuscan}}

\begin{document}
\title{A Large Scale Investigation of Obfuscation Use in Google Play}

% Authors
\author{Dominik Wermke}
\affiliation{%
    \institution{Leibniz University Hannover}
}
\email{wermke@sec.uni-hannover.de}

\author{Nicolas Huaman}
\affiliation{%
    \institution{Leibniz University Hannover}
}
\email{huaman@sec.uni-hannover.de}

\author{Yasemin Acar}
\affiliation{%
    \institution{Leibniz University Hannover}
}
\email{acar@sec.uni-hannover.de}

\author{Brad Reaves}
\affiliation{%
    \institution{North Carolina State University}
}
\email{bgreaves@ncsu.edu}

\author{Patrick Traynor}
\affiliation{%
    \institution{University of Florida}
}
\email{traynor@cise.ufl.edu}

\author{Sascha Fahl}
\affiliation{%
    \institution{Leibniz University Hannover}
}
\email{fahl@sec.uni-hannover.de}

\begin{abstract}
Android applications are frequently plagiarized or repackaged,
and software obfuscation is a recommended protection against these practices.
However, there is very little data on the overall rates of app obfuscation, the
techniques used, or factors that lead to developers to choose to obfuscate
their apps.
In this paper, we present the first comprehensive analysis of the use of and
challenges to software obfuscation in Android applications.
We analyzed 1.7 million free Android apps from Google Play to detect various obfuscation techniques,
finding that only 24.92\% of apps are obfuscated by the developer.
To better understand this rate of obfuscation, we surveyed 308 Google Play
developers about their experiences and attitudes about obfuscation.  We found
that while developers feel that apps in general are at risk of plagiarism, 
they do not fear theft of their own apps.  Developers also self-report difficulties applying obfuscation for
their own apps. To better understand this, we conducted a follow-up study
where the vast majority of \studyvalid{} participants failed to obfuscate a realistic
sample app even while many mistakenly believed they had been successful.
Our findings show that more work is needed to make obfuscation tools more
usable, to educate developers on the risk of their apps being reverse
engineered, their intellectual property stolen, their apps being repackaged
and redistributed as malware and to improve the health of the overall Android
ecosystem.
\end{abstract}

\keywords{Obfuscation, Android, User Study}

% Title in ACM after abstract
\maketitle

\section{Introduction}
\label{sec:intro}
While smartphones have changed society in countless ways, application markets
are perhaps an underappreciated development. These markets have 
enabled the quick and simple distribution of new software, but they have also
enabled numerous studies of application security~\cite{Enck10,Enck11,Enck14},
and provided mechanisms to identify malware on devices before or after
infection~\cite{lock12,chakradeo13}.  Much of this research depends on
automated and or manual software analysis techniques, and these techniques
face challenges in the presence of \textit{software obfuscation}~\cite{collberg2002,hanna2013,zhang2014,linares-vasquez2014,glanz2017}, software transformations designed to frustrate automatic or manual analysis.

Despite the impacts of obfuscation, to-date there is very little \emph{data}
on how Android apps are obfuscated in practice apart from limited or
small-scale studies~\cite{Enck11,Liu15}.  In this paper, we present the first
holistic, comprehensive analysis of the state of the use of software
obfuscation in Android applications.  We begin with a study of obfuscation
usage (and techniques) on over 1.7 million apps collected from Google Play. We follow this with a 
survey of \surveyvalid{} application developers about their
experiences and perceptions of software obfuscation. We conclude with a
development study with \studyvalid{} developers to investigate usability issues with
{ProGuard}, which is by a large margin the most popular obfuscation tool for Android. 
We address three  research  questions:

\par\noindent
\textbf{RQ1:} \emph{How many apps are obfuscated, and what techniques are
used?}   For researchers who
develop automated analysis tools, it is critical to understand what types of
obfuscation are commonly applied -- and at what rate -- so that they can
ensure that they correctly analyze apps.  It is also an important measurement
for the app ecosystem at large. Software obfuscation is a defense against app
repackaging, an abusive practice where entire applications are  cloned and
redistributed to build trojan apps or steal ad revenue.  This practice of
app repackaging is an epidemic threat to the entire
ecosystem:  in recent studies, 86\% of malware samples  collected were
repackaged versions of benign applications~\cite{zhou2012a}, and apps are
repackaged by the thousands~\cite{viennot2014,crussell2015}.  Up to 13\%
of entire third party markets consist of plagiarized, repackaged
apps~\cite{zhou2012b,zhou2013}. Thus, software obfuscation serves to protect not 
just individual apps and developers, but users and the ecosystem at large. 

We find that  roughly 25\% of apps are obfuscated, but that number rises to
50\% for the most popular apps with more than 10 million downloads. This is
high enough that it would have a significant impact on research -- especially
for projects that ignore obfuscated apps~\cite{oltrogge2015,tendulkar2014}.
However, it is also still low enough to indicate that the vast majority of apps are unprotected.

\par\noindent
\textbf{RQ2:} \emph{What are developers' awareness, threat models,
experiences, and attitudes about obfuscation?}
These factors provide insight into the root causes of  the low rates of
obfuscation in Android. We examine whether developers are aware of
obfuscation, whether they report to have attempted or successfully used
obfuscation, which tools they have used,  and whether they found the tools
were sufficiently easy to use.
We find that while developers are aware of the benefits of obfuscating
their  apps  on  a  theoretical  level,  a  perceived  negligible personal
impact and the time-consuming use of obfuscation tools in real world applications is a
large deterrent to using obfuscation.

\par\noindent
\textbf{RQ3:} \emph{How usable is the leading obfuscation tool?}
Our developer survey also found that 35\% of our participants reported 
difficulty obfuscating their apps, while over 61\% --- more than double
the Play market average --- claim to obfuscate their apps. To better understand
this paradox, we asked \studyvalid{} developers to obfuscate two sample apps.
We found that while most developers successfully managed to complete a simple
obfuscation task, 78\% failed to correctly use {ProGuard} in a more complex and realistic scenario. Moreover, 38\% mistakenly believed they had successfully obfuscated their app. This highlights that even when developers attempt to use obfuscation,
tool usability likely has a negative impact on its effectiveness.

We conclude our paper with a discussion of lessons
learned and future directions  for improving this state of affairs in
Section~\ref{sec:discussion}. 
We acknowledge that software obfuscation is not a ``silver bullet'' that
defends against all reverse engineering, but previous work shows that even
simple forms of obfuscation (like identifier renaming) significantly increase
the effort required to successfully reverse engineer
software~\cite{ceccato2009,ceccato2014}. Additionally, the significant
challenges obfuscation presents researchers (as shown in prior work~\cite{hanna2013,zhang2014,linares-vasquez2014,glanz2017})
make this topic worthy of study. We also note our focus is on obfuscation
practices used by legitimate applications; we leave the topic of obfuscation
of malware for future work.

We note that the implications of this study go
far beyond the Android ecosystem. In contrast to other secure practices with a variety
of costs and trade offs, software obfuscation is in an ideal position for
adoption: {ProGuard} is one of the very few secure development tools in
existence that is free, already available in the IDE of most developers, and
can \emph{automatically enhance} security while simultaneously
\emph{improving performance}. Understanding why developers do or do not use
such an ideal tool has broad implications both for the development of better
developer support and as a measure of barriers 
to a more security-conscientious software
development community.

\section{Android Obfuscation Techniques \& Tools}
\label{sec:background}

\label{subsec:background}
% Usage: \rot[angle=90][width=0.2em]{text}
\NewDocumentCommand{\rot}{O{90} m}{\makebox[0.2cm][l]{\rotatebox{#1}{#2}}}%

\begin{table}
    \centering
    \begin{threeparttable}
    \setlength{\tabcolsep}{2pt}
    \footnotesize
    \begin{tabular}{%
        l@{\hspace{8\tabcolsep}}l@{\hspace{4\tabcolsep}} % Tool, Type
        ccccccc@{\hspace{4\tabcolsep}} % Obfuscation
        cc % Encryption
        ccc % Other features
    } 
        \toprule
        &&\multicolumn{7}{c}{Obfuscation}&\multicolumn{5}{c}{Other}\\
        \cmidrule(r{8pt}){3-9}\cmidrule{10-14}
        \textbf{Name}& License &
        \rot{Package name}&\rot{Class name}&\rot{Method name}&\rot{Field name}&\rot{Overloading}&\rot{Debug Data}&\rot{Annotations}&
        \rot{String Enc.}&\rot{Class Enc.}&\rot{Optimization}&\rot{Minimization}&\rot{Watermarking}\\
        \midrule
        Allatori\tnote{1,$\dagger$} & \$290 & \yes & \yes & \yes & \yes & \yes & \yes & \no &
        \yes & \no & \yes & \yes & \yes\\
        %APKProtect & Free & \multicolumn{11}{c}{Not maintained since 2013}\\
        DashO\tnote{$\dagger$}& On request & \yes & \yes & \yes & \yes & \yes & \yes & \yes &
        \yes & \no & \yes & \no & \yes\\
        DexGuard\tnote{2,$\dagger$} & On request & \yes & \yes & \yes & \yes & \yes & \yes & \yes &
        \yes & \yes & \yes & \yes & \no\\
        DexProtector & \$800 & \yes & \yes & \yes & \yes & \no & \no & \no &
        \yes & \yes & \no & \no & \no\\
        GuardIT & On request & \yes & \yes & \yes & \yes & \yes & \yes & \no &
        \yes & \yes & \no & \no & \no\\
        Jack\tnote{2,$\dagger$} & Free  & \yes & \yes & \yes & \yes & \no & \yes & \yes &
        \no & \no & \yes & \yes & \no\\
        ProGuard\tnote{$\dagger$} & Free & \yes & \yes & \yes & \yes & \yes & \yes & \yes &
        \no & \no & \yes & \yes & \no\\
        ReDex\tnote{2,$\dagger$} & Free & \yes & \yes & \yes & \yes & \yes & \yes & \yes &
        \no & \no & \yes & \yes & \no\\
        yGuard\tnote{$\dagger$} & Free & \yes & \yes & \yes & \yes & \yes & \yes & \no &
        \no & \no & \no & \yes & \no\\
        \bottomrule
    \end{tabular}
    \begin{tablenotes}
        \footnotesize
        \item[1]Multiple obfuscation patterns, default can be detected
        \item[2]Mirrors ProGuard's obfuscation with same configuration format
        \item[$\dagger$] Obfuscation features (partially) detected by \toolname{}
    \end{tablenotes}
     \caption[Popular obfuscation software.]{Selected features of popular obfuscation software for the Android environment.}
    \label{tbl:bg:tools}
    \end{threeparttable}
\end{table}

Obfuscation tools for the Android ecosystem cover a wide range of prices and features.
Available tools range from free, open-source obfuscation solutions providing only basic obfuscation features such as {ProGuard},
up to premium obfuscation environments with high monthly per-developer-licensing fees such as {DexGuard} (cf.\ Table~\ref{tbl:bg:tools}).

The free {ProGuard} enjoys preferential treatment in the Android ecosystem. It is included with the Android SDK and supported by the official Android Studio IDE.
In addition, other obfuscation tools inherit most of their obfuscation functionality from {ProGuard}; the now deprecated alternative tool chain Jack is configured by {ProGuard} configuration files and provides {ProGuard}'s obfuscation with reduced options.
Similarly, {ReDex} accepts {ProGuard}'s configuration files and mirrors the renaming functionality closely. {DexGuard} is a commercial {ProGuard} extension and utilizes name obfuscation with the same basic functionality as {ProGuard}, but with extended features and symbol space.

\begin{lstlisting}[caption={Example configuration to enable {ProGuard} in the Gradle build system.
Configured in the \emph{build.gradle} file of an {Android Studio} project.},label={lst:build},float]
android{
   buildTypes {
       release {
           minifyEnabled true
           proguardFiles 'proguard-rules.pro'
}}}
\end{lstlisting}

\begin{lstlisting}[caption={Example {ProGuard} configuration. Configuration path is set in the build system, e.g.\ in a \emph{gradle.build} file.},label={lst:proguard},float]
-optimizationpasses 5

-dontusemixedcaseclassnames
-overloadaggressively
-printmapping mapping.txt

-keep public class * extends project.Interface
-dontwarn project.example.**
\end{lstlisting}

{ProGuard} has been integrated with the Android Software Development Kit (SDK) since August 2009 and can be activated in the build setup of a project.
The ``\texttt{minifyEnabled}'' option activates {ProGuard} obfuscation for the release build of an app.
Additional configuration files can be specified with the ``\texttt{proguardFiles}'' option.

In the {ProGuard} configuration file, different program options are activated/deactivated by setting a number of flags that are relevant to later presented results (cf.\ Listing~\ref{lst:proguard}).
Some processing steps of {ProGuard} can be completely disabled with flags such as ``\texttt{-keep}''.

Android obfuscation techniques are applied to different components of an application:
\begin{description}
\item[Name obfuscation.]{%

\textbf{Package}, \textbf{class}, \textbf{method}, and \textbf{field} names are commonly obfuscated by replacing their original values with meaningless labels.

By default, {ProGuard} implements name obfuscation by generating name replacements using characters from the \texttt{[a-zA-Z]} alphabet. Obfuscated names are generated by iterating through the alphabet resulting in the following pattern: \texttt{a}, \texttt{b}, \ldots, \texttt{z}, \texttt{A}, \ldots, \texttt{Z}, \texttt{aa}, \texttt{ab}, \ldots, \texttt{zz}.
However, users can add their own word lists to the renaming alphabet.
{Allatori} and {DexGuard} build on {ProGuard's} name obfuscation alphabet and add reserved Windows keywords (``AUX'', ``NUL'').

\begin{center}
    \noindent
    \begin{minipage}{0.9\columnwidth}
        \begin{minipage}[t]{0.45\columnwidth}
            \begin{lstlisting}
public class Matrix {
	private int M;
	public Matrix(int M);
}
            \end{lstlisting}
        \end{minipage}\hfill
        \begin{minipage}[t]{0.45\columnwidth}
            \begin{lstlisting}
public class a {
	private int a;
	public a(int b);
}
        \end{lstlisting}
        \end{minipage}
    \end{minipage}
\end{center}
}

\item[Name overloading.]{%
Exploiting the overloading feature of the Java programming language, obfuscation tools commonly assign the same name to methods with different signatures (i.\,e. different list of method argument types). In addition to use the same for different methods, method parameters are renamed using name obfuscation techniques.

\begin{center}
    \noindent
    \begin{minipage}{0.9\columnwidth}
        \begin{minipage}[t]{0.45\columnwidth}
            \begin{lstlisting}
public class Matrix {
	public Matrix(int M);
	public Update(double D);
}
            \end{lstlisting}
        \end{minipage}\hfill
        \begin{minipage}[t]{0.45\columnwidth}
            \begin{lstlisting}
public class a {
	public a(int a);
	public a(double a);
}
            \end{lstlisting}
        \end{minipage}
    \end{minipage}
\end{center}
}

\item[Debug data obfuscation.]{%
The removal of debug information printed in stack traces such as line numbers or method names complicates the reverse engineering of code structures by intentionally-caused error stack traces.
Obfuscation tools generally include means to reverse the information removal to allow for debugging of the app during development.
}

\item[Annotation obfuscation.]{%
Another feature related to the removal of information strips annotations from classes and methods.
This includes annotations such as ``Inner Class'' for inner classes or ``Throws'' for methods that contain throw statements.
Annotations allow for the retrieval of additional functional context from encountered classes.
Similar to debug information, the removal of class file annotation and the removal of class source file information complicates the reverse engineering of code structures by tracing class attributes.
}

\item[String encryption.]{%
Strings can be encrypted to hide information. A trade-off has to be made between encryption strength and performance impact by decryption. The decrypter has to be provided in the program, making encryption unsuitable to hide sensitive information.
Strings are encrypted to deter simple string searches over the code base and hide information about the program flow.
}

\item[DEX file encryption.]{%
The \emph{classes.dex} file can be encrypted to avoid detection by decompilers and to increase the difficulty of decompilation.
Decryption of encrypted classes at run time can cause large performance impacts. 
}
\end{description}

\paragraph{Complications for Obfuscation}
While the previous section has discussed a number of techniques for transforming software,
configuring obfuscation tools for Android is more complicated than merely choosing from the available features.
In fact, there are a number of complicating situations that make it difficult or impossible to obfuscate certain pieces of code,
and if that code happens to be obfuscated the app can no longer function.
These situations for partial obfuscation include classes that need to be accessible from an outside context:
the names and class names of native methods and similarly classes that extend native Android classes such as activities,
services or content providers should remain unobfuscated in most cases so that the library/system can invoke callbacks.

\section{Detecting {ProGuard} Obfuscation}
\label{sec:obfuscan}
To answer our research question on \emph{how many apps are obfuscated, and what techniques are
used}, we built a tool we call \toolname{} to conduct a large scale measurement study of obfuscation practices.
\toolname{} is able to detect a number of obfuscation features in compiled Android binaries.
In particular, \toolname{} is able to detect all of {ProGuard}'s obfuscation features and many features of other obfuscation tools (as shown in Table~\ref{tbl:bg:tools}).

\paragraph{How \toolname{} Works}
\toolname{} takes an Android binary as input and analyzes certain parts of the binary to detect specific obfuscation features and outputs the list of all detected features.
\toolname{} analyzes package, class, method and field names to detect name obfuscation.
To detect method name overloading, \toolname{} analyzes the distribution of obfuscated method names for duplicates and relies on the content of debug entries to detect debug information removal.
Annotation removal is detected by analyzing an app binary's for the removal of corresponding class attribute fields.
To detect further obfuscation features, \toolname{} relies on the \textit{classes.dex} file format and specific function calls (see below).

\paragraph{Feature Detection}
\label{subsec:tool_obf_detection}

\toolname{} implements a number of heuristics to detect obfuscation features. To ensure accuracy, many of these are developed deterministically and directly from the source code of {ProGuard}.

For name obfuscation, \toolname{} detects both lower- and upper-case obfuscated names by simulating the obfuscation process of {ProGuard} and comparing the generated names to the actual names encountered on the app, package, or class level.
\toolname{} also considers possible flags such as the usage of mixed-case characters if corresponding strings are detected in the scope. 
Finally, \toolname{} also looks for instances where tools replace class names with restricted keywords in the Windows operating system utilized by {DexGuard} and some Allatori configurations.
To detect method name overloading, \toolname{} investigates names that follow the obfuscation pattern and occur multiple times on the same class level.
\toolname{} detects missing debug information by parsing and storing the entries of the Java \emph{LineNumberTable} which maps bytecode instruction to source code line numbers.
Similarly, the removal of the source file data from classes removes information about the source file where the class (or at least its majority) is defined.
\toolname{} detects this feature by directly accessing the source file attribute of classes and storing the string content of the attribute.
Removal of annotations is detected by \toolname{} by directly accessing and storing the attribute field of classes.

\paragraph{Other Tools}
Although we built \toolname{} with a focus on detecting the use of {ProGuard},
it is able to detect apps that were obfuscated with other tools (cf. Table~\ref{tbl:obfsc:eval:obfs_email}).
\toolname{} is able to detect apps that were obfuscated using {ReDex}, {Jack} and {DexGuard} name obfuscation using \toolname{}'s name obfuscation detection feature since all three tools use name obfuscation patterns that are identical with {ProGuard}'s name obfuscation. Additionally, \toolname{} is able to detect {DexGuard}'s more advanced removal of debug line numbers and annotations obfuscation features. We extended \toolname{}'s name obfuscation detection feature to also cover the name obfuscation patterns implemented by {yGuard} and {DashQ}. To be able to detect {Allatori}'s non-alphanumeric name obfuscation scheme, we extended \toolname{} and added detection support for restricted Windows keywords such as ``AUX'' or ``NUL''.

\paragraph{Evaluation}
We implemented \toolname{} in Python and evaluated its efficacy by conducting a lab experiment using 100 real Android applications randomly selected  from the {F-Droid} open source app market. 
We compiled two different versions of each sample app: One version did not use any means of obfuscation and one version that had {ProGuard}'s name obfuscation for all application scopes, method name overloading, debug information removal, annotation removal, and source file removal enabled. Additionally, we acquired and tested 26 apps obfuscated with {DexGuard}, an expensive commercial tool, correctly identifying obfuscation in all 26.

\begin{table}
	\centering
	\begin{tabular}{lrrrrrrrr}
		\toprule
		Feature                     & TP    & TN    & FP    & FN    & MCC\\
		\midrule
		Class name obfuscation      & 98   & 100   & 0     & 2     & 0.980\\
		Method name obfuscation     & 99   & 100   & 0     & 1     & 0.990\\
		Field name obfuscation      & 100   & 92   & 8    & 0     &\textbf{0.923}\\
		Method name overloading     & 99   & 100   & 0     & 1     & 0.990\\
        \midrule
		Debug information removed   & 100   & 100   & 0     & 0     & 1.000\\
		Annotations removed         & 100   & 88   & 12    & 0     &\textbf{0.886}\\
		Source files removed        & 100   & 100   & 0     & 0     & 1.000\\
		\bottomrule
	\end{tabular}
	\caption{Performance of \toolname{} for sample set of 200 {APKs}.
	Shown are true positive (TP), true negative (TN), false positive (FP), false negative (FN) predictions,
	and Matthews correlation coefficient (MCC).}\label{tbl:obfsc:det:heu_perf}
\end{table}

\toolname{} correctly identifies nearly all obfuscation features the 200 {APK}s dataset with a low false-positive rate and a high correlation coefficient (cf.\ Table~\ref{tbl:obfsc:det:heu_perf}). We manually investigated false positives and false negatives. \toolname{} falsely detected few class and method names as not obfuscated. In these cases, structures of the app were exempt from obfuscation,
e.g.\ due to classes being marked as an interface. 
The false positive rate for field names is slightly higher than for other features.
This is because {ProGuard} uses short strings for names (e.g., a and b) that are sometimes used as variables in unobfuscated apps. 
\toolname{} had no false positives for the debug information and source files removal feature.
However, it falsely detected 12 apps as using the annotations removal feature. These false positives affect apps that do not use the code characteristics that are compiled to annotations (like inner classes). 

\paragraph{Limitations}
There are several obfuscation features that \toolname{} does not measure. Since \toolname{} focuses on the detection of the benign application of obfuscation, we do not look for packers or other techniques specifically used by malware. We excluded the heuristics for resource name and content obfuscation from our large scale measurement study for performance reasons. We evaluated a test set of 1,000 random apps from Google Play and could not find a single app using these features. Additionally, we did not implement class and string encryption detection. Both are advanced features and {DexGuard}, {DexProtector}, or {GuardIT} provide them as extensions to the more basic name obfuscation features. Finally, \toolname{} focuses on the detection of name obfuscation as implement by common tools. These heuristics conservatively estimate the prevalence of obfuscation at the cost of missing the use of name obfuscation algorithms by less popular tools.
However, because  \toolname{} reliably detects the removal of debugging information, we believe that this estimates a strong upper bound of the potential uses of other tools besides ProGuard-related tools.

\toolname{}'s annotation removal detection looks for application packages that do not include annotations. However, this heuristic might mislabel unobfuscated apps that naturally do not use annotations. Since it is hard to estimate in how many cases this specific heuristic reports false positives, we excluded it from our large scale measurement study in Section~\ref{sec:second_case}.

To test the efficacy of Obfuscan, we used apps from F-Droid rather than Google Play because we needed access to source code; while there is a chance that F-Droid apps differ from Google Play apps, this methodology was better than alternatives like writing self-generated apps.

\section{\toolname{} Analysis Results}
\label{sec:second_case}
\begin{figure*}
    \centering
    \includegraphics[width=0.8\textwidth]{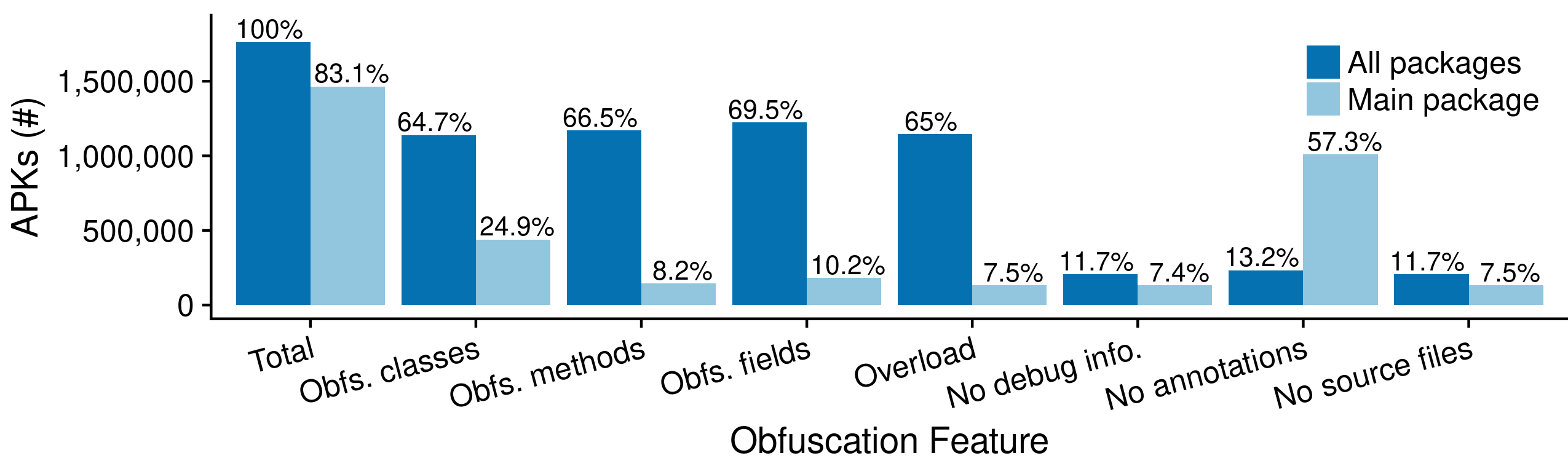}
	\caption[Obfuscation numbers of app structures]{%
	    Comparison of obfuscation for different app structures including all packages and main package only.
	    Only apps with an identifiable main package are included in the corresponding category.
	    Overall obfuscation of apps considering all packages is increased due to library obfuscation.}
	\label{fig:obfsc:eval:main1}
\end{figure*}

We performed a large-scale analysis of \allanalyzedapps{} current free Android apps from Google Play to investigate the real-world use of the {ProGuard} family of obfuscation tools. To the best of our knowledge, this is the largest obfuscation detection analysis to-date for Android applications.
Of those applications, \toolname{} detected the renaming obfuscation pattern implemented by the {ProGuard} family of obfuscation tools (cf.\ Section~\ref{sec:background}) in 1,137,228 (64.51\%) apps.

However, a large percentage of apps were not intentionally obfuscated by the
original developer, but contained third-party libraries that used obfuscation.
While some libraries are distributed pre-obfuscated, others ship with {ProGuard}
configuration files to configure obfuscation. The fact that libraries may be obfuscated, but main application code
non-obfuscated, is an important distinction for understanding the use of
obfuscation throughout the Android ecosystem. In particular, the presence of
an obfuscated library does not indicate that core application components are
actually being obfuscated.

\paragraph{Obfuscation in Libraries}
To get a better overview over the included libraries in the Android ecosystem, we investigated the names of Android packages in all apps.
Android packages follow Java naming conventions, allowing for the identification of larger scopes
(e.g.\ the \emph{com.google.ads.interactivemedia.v3.api} package can be traced to the \emph{com.google.ads.*} scope).
Analyzing the scope distribution of obfuscated packages across the apps, it emerges that most of the external library obfuscation
stems from a few, popular library frameworks (cf.\ Table~\ref{tbl:obfsc:eval:most_lib}).

\begin{table}
	\centering
	\tabcolsep=0.08cm
	\begin{tabular}{lr>{\bfseries}r}
		\toprule
		Scope                       &   Packages    &   Unique APKs\\
		\midrule
		com.google.ads.*            &   1,919,976   &   681,102\\
		com.google.android.gms.*    &   24,095,920  &   651,952\\
		android.support.v4.*        &   1,811,806   &   192,497\\
		com.unity3d.*               &   432,856     &   152,668\\
		org.fmod.*                  &   135,524     &   135,524\\
		android.support.v7.*        &   992,843     &   117,680\\
		com.facebook.*              &   1,309,276   &   106,178\\
		com.startapp.*              &   2,234,609   &   88,242\\
		com.chartboost.*            &   491,612     &   87,781\\
		com.pollfish.*              &   537,046     &   44,851\\
		\bottomrule
	\end{tabular}
	\caption{Most prevalent obfuscated libraries by total number of packages and number of {APK}s containing libraries of the scope.
	The scope of the libraries is defined by their package name structure.}\label{tbl:obfsc:eval:most_lib}
\end{table}

Examples include the Google Ad framework used in the monetization of apps and
the Google Mobile Service ({GMS}) framework for interfacing with Google services such as authentication or search.
Commonly obfuscated frameworks not related to Google include the Facebook framework for integrating Facebook access into apps and the {FMOD} library for audio playback.
The Google frameworks for ads and services are commonly used in apps for basic features,
adding obfuscated packages to a large number of apps.
The presence of these very popular libraries explains why many
applications are shown to be obfuscated when examined on an overall package
basis, but so few main packages are obfuscated.

\paragraph{Application Obfuscation Rates}
While identifying popular libraries is easy to do manually, separating
developer code (which may be similar among several apps by the same
development team) from less common libraries is far more difficult~\cite{backes2016}.
To distinguish between apps that are obfuscated by their developer and apps
that simply include obfuscated libraries, we also analyze the obfuscation used
by the declared main package of the application
(This distinction of main package vs.\ other packages was also performed by Linares-V{\'a}squez et al.~\cite{linares-vasquez2014}).
The main package is used as the universal identifier of the application (e.g. \emph{com.google.maps}) and is
necessarily implemented by the developer, so a choice to obfuscate the main
application strongly indicates a choice to obfuscate at least some 
(if not all) of the original application code. 

Our main package analysis found that only 24.92\% of apps (\obfuscatedapps{} apps) are
intentionally obfuscated by the developer. In other words, \emph{ the vast majority of apps ---
representing millions of man-hours of development --- are not protected using
{ProGuard} as recommended for use in the official Android developer documentation~\cite{android-api}.}

\paragraph{Obfuscation Feature Popularity}
\toolname{} provides the ability to examine use of individual {ProGuard}
obfuscation features, and the use of these features for both entire
applications and main packages only is shown in Figure~\ref{fig:obfsc:eval:main1}. 
The ``all package'' category is measured as the number of apps containing any package with the obfuscation feature.
This includes all libraries and the declared main package.
The ``main package'' category is the number of apps with the obfuscation feature considering only the app's main package.
We note that percentages of features used in the main package results are 
only among those apps with code in the main package.

We see first that class name obfuscation is the most popular feature, with
64.7\% of all packages and 24.9\% of main packages using it. Looking at other
features shows a marked difference in feature use between libraries and main
packages. While features that obfuscate method names, field names, and exploit
function name overloading are used about as often as class name
obfuscation in the all package analysis, they are infrequently used in main
packages. One explanation is that library developers have a greater incentive
to protect proprietary or sensitive internal {API}s.

In addition to name obfuscation features, we also investigated the information
removal features of {ProGuard} for the main package and all packages.  As
shown in the validation dataset, these features are generally a weaker
indicator of obfuscation because their presence depends on characteristics of
the code base.
For example, the large percentage of main packages without annotations stems from basic code without inner classes,
exceptions, or functionality that would require annotations.
For all packages, percentages for these features are lower than the name obfuscation features.
Library developers may omit these obfuscation features from their configurations to enable debugging by end developers. 

Overall, our findings indicate that the vast majority of app developers
do not obfuscate their core code, and that even when they do they do not use 
all of the available obfuscation features. These results might indicate that
developers either only obfuscate critical parts of their application or do not
understand the entire concept of obfuscation.

\paragraph{Non-Proguard Obfuscation}
While \toolname{} comprehensively covers features used by ProGuard, it also provides information about other forms of obfuscation.
First, apps that do not contain debug info or source files are likely obfuscated, and so looking for those characteristics provides an upper bound on the number of apps in our dataset that are obfuscated by any non-ProGuard tool.
As shown in  Figure~\ref{fig:obfsc:eval:main1}, we find that between 7.4 and 7.5\% of apps in our data have these features for the main package, while between 11.7 and 13.2\% of apps have these features for any class in the application.
Additionally, we found 2,799 (0.16\%) apps that use the advanced obfuscation feature of replacing class names with restricted keywords of the Windows operating system (e.g. ``AUX'', utilized by {DexGuard} and some Allatori configurations).
By analyzing \emph{classes.dex} files, we found 794 (0.05\%) apps that were obfuscated with {DexProtector} and 207 (0.01\%) apps obfuscated with Bangcle. Ultimately, these results together allow us to conclude that {ProGuard} is far more popular than any other obfuscation tool. This is because the classes using {ProGuard}-style name obfuscation greatly outnumber the scrubbed debugging or source files, which provide an upper bound on all other obfuscation tools.

\subsection{Obfuscation Trends}

By comparing our obfuscation findings with Google Play metadata for all apps that we analyze, we
can develop further insights into the use of obfuscation in Android. 
In this subsection, we consider an app "obfuscated" if classname obfuscation
is used, as this is the most common obfuscation feature supported by most obfuscation tools.
As in the previous subsection,
we distinguish between ``all packages'' and ``main packages'' for our
analysis.
We investigate following trends in app obfuscation:
main package obfuscation rate in relation to download numbers;
average main package obfuscation by number of apps per Google Play account; and obfuscation by app update date.

\begin{figure}[ht!]
    \centering
    \includegraphics[width=\columnwidth]{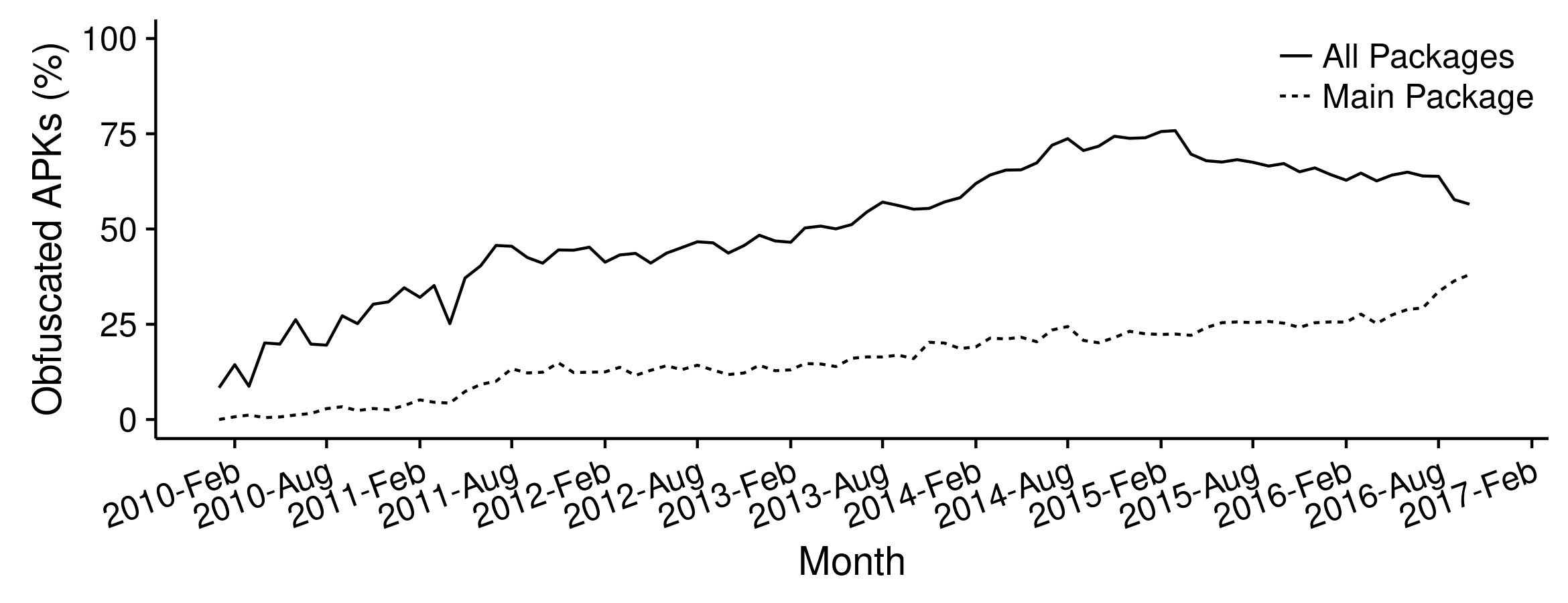}
	\caption[Percentage of obfuscated apps by update month]{Comparison between the percentage of all obfuscated apps and the percentage of apps with obfuscated main package among the apps updated
	each month. Update dates are gathered from Google Play metadata and categorized to months. Percentage of apps with obfuscated main package increases for more recent update dates.
	\label{fig:obfsc:eval:date_dist}}
\end{figure}

\begin{description}
\item[App Popularity:]{%
\begin{table}
	\centering
	\tabcolsep=0.11cm
	\begin{tabular}{rr>{\bfseries}r}
		\toprule
		Download Counts & Total Apps    & Obfs. Main Package\\
		\midrule
		0+              & 115,683       & 27.30\%\\
		10+             & 343,652       & 26.34\%\\
		100+            & 499,018       & 24.74\%\\
		1,000+          & 383,046       & 24.13\%\\
		10,000+         & 234,213       & 23.95\%\\
		100,000+        & 80,302        & 25.50\%\\
		1,000,000+      & 16,335        & 29.15\%\\
		10,000,000+     & 1940          & 36.80\%\\
		100,000,000+    & 160           & 50.00\%\\
		\bottomrule
	\end{tabular}
	\caption[Distribution of main package obfuscation for different download counts.]{%
	    Distribution of main package obfuscation for different download counts.
	    More popular apps have a higher rate of main package obfuscation.}
	\label{tbl:obfsc:eval:download_main}
\end{table}
Google Play apps range from rarely downloaded one-off weekend
projects to popular and complex apps with dozens of developers and millions of installs.
Hence, different apps will have different incentives to obfuscate their code.
We hypothesized that popular apps would be more likely to obfuscate their code
as these apps are often more sophisticated and complex, but also face the
greatest risks of plagiarism. To test this hypothesis, we compared the rates
of obfuscation for each download count category reported by the Google Play
market.

Table~\ref{tbl:obfsc:eval:download_main} shows these results. We find that
most apps --- the 98.9\% (1,655,914 apps) of apps with less than 1 million downloads
--- are obfuscated at roughly the same rate, ranging from 23.9\% -- 27.3\%.
As download counts increase further, we see an increase in obfuscation in the
most downloaded apps from 29.15\% of apps with more than one million
downloads to 50.0\% of apps with more than 100 million downloads. While this
does confirm our initial expectation, we were surprised that even the most
popular apps are only obfuscated on average half of the time.
}

\item[Obfuscation by Google Play account:]{%
\begin{table}
	\centering
	\tabcolsep=0.11cm
	\begin{tabular}{rr>{\bfseries}r}
		\toprule
		Apps per Account  & Unique Accounts & Avg. Obfs. of MP\\
		\midrule
		1       & 311,908   & 21.83\%\\
		2+      & 155,220   & 21.24\%\\
		10+     & 27,397    & 26.50\%\\
		100+    & 642       & 34.37\%\\
		250+    & 112       & 35.29\%\\
		500+    & 36        & 68.41\%\\
		\bottomrule
	\end{tabular}
	\caption[Average main package obfuscation by Google Play account]{Average main package obfuscation for number of apps by Google Play account.
	Accounts with more apps have a higher average rate of main package obfuscation.}\label{tbl:obfsc:eval:obfs_email}
\end{table}
Similar to app popularity, we also investigated if the number of published apps
per Google Play account plays a role in the decision to obfuscate apps.
Our hypothesis was that accounts with more submitted apps either belong to experienced
developers or even companies specialized in app development and that apps from these
accounts would show a higher obfuscation rate either due to a higher awareness or even
previous experience of intellectual property theft or due to a higher perceived
investment.

Table~\ref{tbl:obfsc:eval:obfs_email} shows the results.
We find that apps from accounts with less than 100 apps have roughly the same
average obfuscation rate between 21.8\% -- 26.5\%.
For accounts with 100 or more submitted apps this increases to about 35\%
and even to 68.4\% for accounts with 500 and more apps.
This increase in average app obfuscation seems to confirm our hypothesis
that experienced developers or specialized companies with a large number of submitted
apps use obfuscation more often.
A likely explanation for this could be that more experienced developers and companies want to protect their intellectual property further.
This could be the results from previous experiences of intellectual property theft,
or the result of placing a higher value on their apps,
as they are likely an important source of income for professional developers and specialized app companies.
}

\item[Update Date:]{%
Figure~\ref{fig:obfsc:eval:date_dist} shows how all package and main package
obfuscation rates vary when compared to the month of their most recent update;
recent updates on average imply frequent maintenance of apps~\cite{potharaju2017}.\footnote{Unfortunately, our data collection only allowed us to collect the most recent data on an application, preventing us from getting ground truth on the changes in obfuscation of individual apps over time.}
{ProGuard} is distributed with the Android {SDK} starting August 2009.
The base {ProGuard} name obfuscation algorithm remained functionally unchanged,
allowing \toolname{} to detect obfuscation for all included apps over the study period.

The figure shows a clear upward trend for both all packages and main packages,
though as seen previously the overall obfuscation rate for all packages is
much greater than main package obfuscation rate. More recently updated apps are
more likely to be obfuscated as well. This could be indicative of greater
developer sophistication or greater investment in terms of development time and
intellectual property. In any case, it is clear that more recently updated apps are 
more likely to be obfuscated yet are still obfuscated at a low rate.
}
\end{description}

\section{Developer Survey}
\label{sec:survey}
To understand \emph{what developers' awareness, threat models,
experiences, and attitudes about obfuscation are}, we conducted an online survey of Android developers covering their obfuscation experience, the tools they use and their general knowledge and risk assessment concerning obfuscation and reverse engineering. We asked them whether they had heard of obfuscation, whether they knew what it was, whether they had ever used it or decided against using it, and why. Additionally, we measured their awareness of ``repackaging'', ``reverse engineering'', ``software plagiarism'', and ``obfuscation''. We asked how strongly they feel that apps in general and their own apps in particular are threatened by the first three concepts. We followed this up with a set of general questions about their Android development practices.\footnote{Full questionnaire included in the appendix}
In this section, we briefly discuss the design of this survey as well as the results. The online study was approved by the Institutional Review Boards of both involved universities (See Appendix~\ref{app:ethics} for more details).

\begin{figure}
    \centering
    \includegraphics[width=\columnwidth]{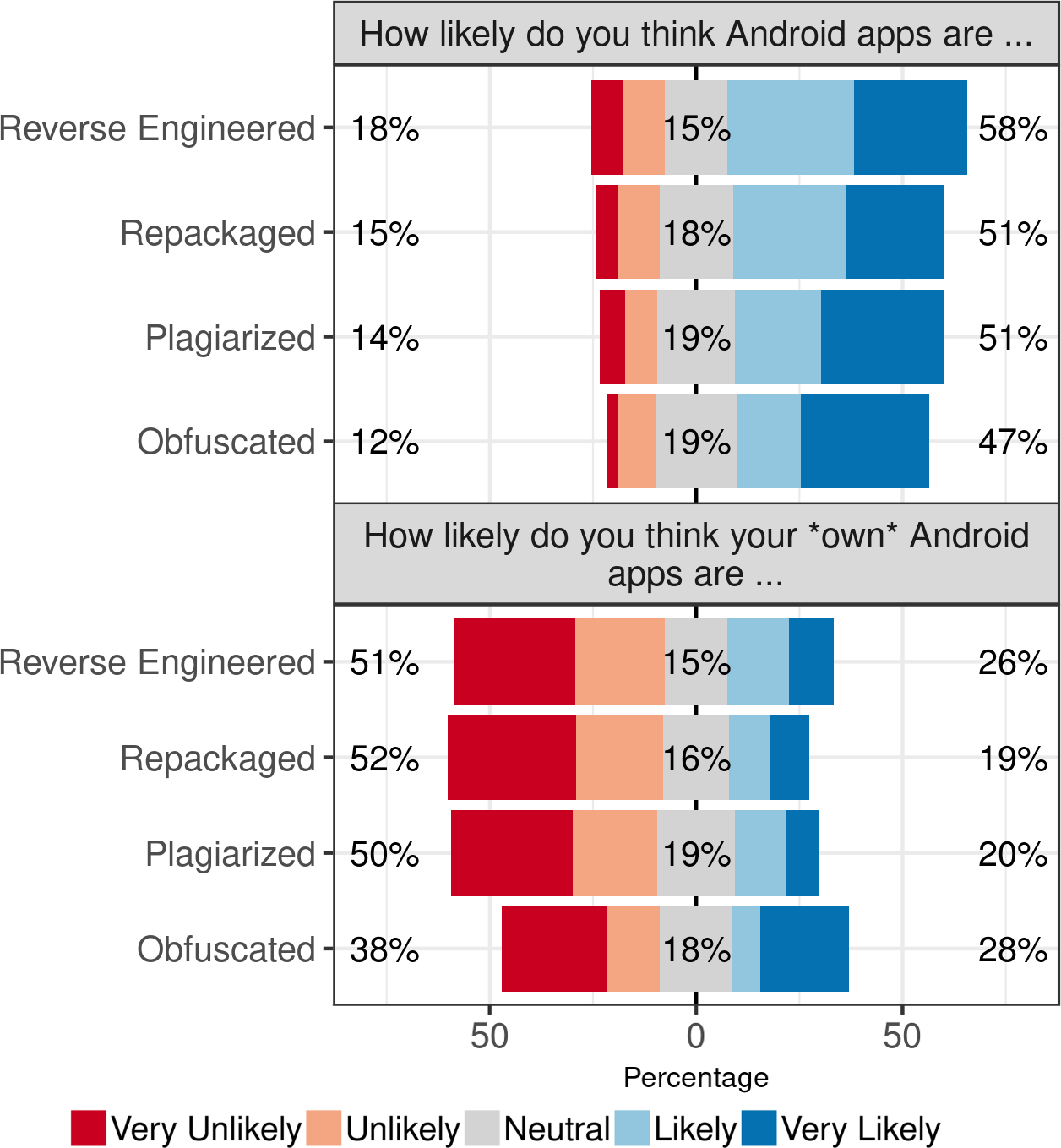}
    \caption[Online Questionnaire Likert.]{Answer distribution of the online questionnaire as Likert plots.
    ``Don't know'' answers are omitted.}
    \label{fig:study:likert_combined}
\end{figure}

Depending on their answers, we asked up to three free text questions, the results of which we analyzed by using open coding them with two researchers, developing an initial codebook and refining it iteratively, using it independently on the answers and resolving all conflicts with the help of a third researcher~\cite{charmaz2014constructing}.

\label{sec:survey:qual}
\begin{description}
\item[Recruiting.]{%

\begin{figure}
    \centering
    \includegraphics[width=\columnwidth]{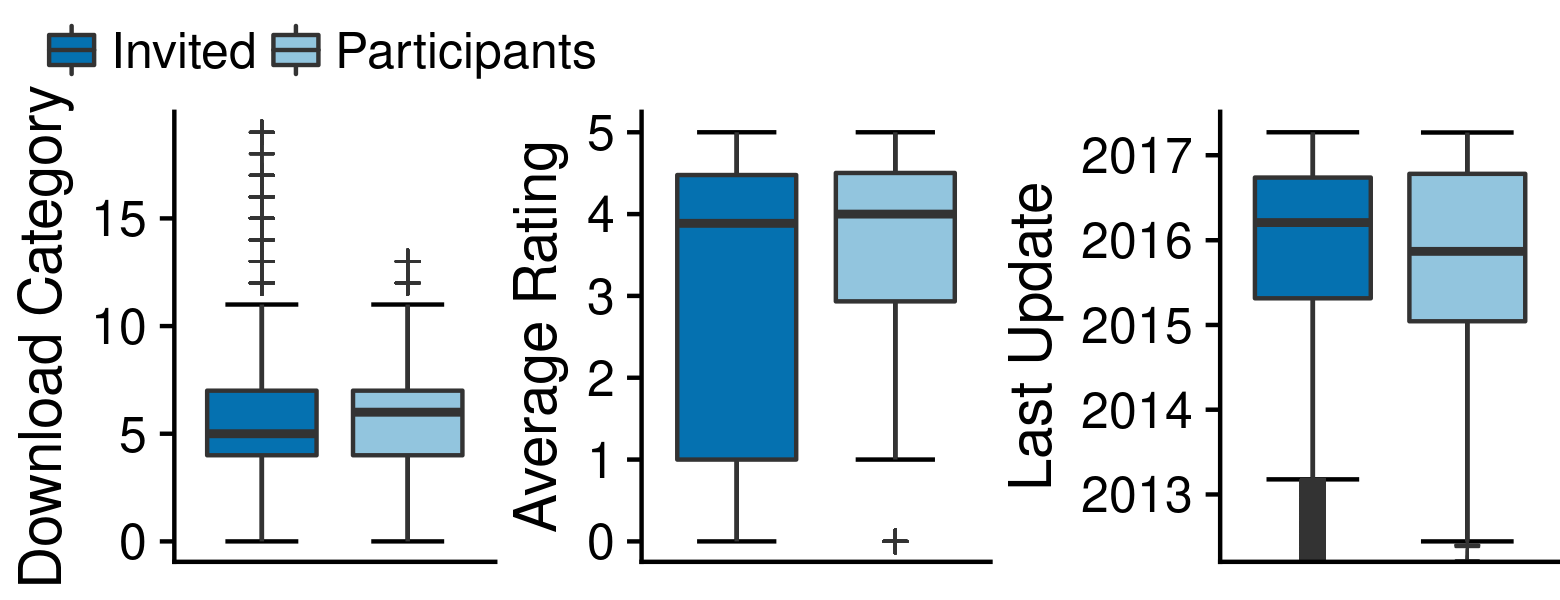}
    \caption[Comparison of invited app metadata vs.\ survey app metadata]{App metadata associated with invited email addresses compared to metadata from our participants.}
    \label{fig:survey:sampling:bp}
\end{figure}

We collected a random sample of \surveyinvites{} email addresses of Android application developers listed in Google Play. We emailed these developers, introducing ourselves and asking them to take our online survey. A total of 561 people clicked on the link to our survey, visited our website and agreed to the study's consent form. Of these 561, 186 dropped out before answering the first question; another 67 participants were removed for dropping out later during the survey or providing answers that were nonsensical, profane, or not in English. Results for our survey are presented for the remaining \surveyvalid{} valid participants.

To determine if our sample was representative of ``typical'' Google Play developers, we compared metadata of 3,159 Android apps associated with Google Play accounts from our survey participants with the metadata of 1.1M free and paid applications associated with the \surveyinvites{} email addresses to which we sent survey invitations (shown in Figure~\ref{fig:survey:sampling:bp}).

We found a close resemblance in download counts per app (mean invited: 5.75, mean participated: 5.89, category 5 corresponds to 100--500 downloads, category 6 to 500--1,000 downloads), the average user rating (mean invited: 3.07, mean participated: 3.29) and the date of the last update as a measure of app age and long-term developer support (mean invited: 2015-11-18, mean participated: 2015-09-01). These similarities suggest that the developers who opted into our survey strongly resemble the random sample of Google Play and therefore the whole Google Play Android developer population.
}

\item[Obfuscation Experience.]{%
We found that the majority (241, 78\%) of our participants had heard of software obfuscation in general, while 210 (68\%) knew about obfuscation techniques for Android in particular. 187 (61\%) had considered obfuscating one or multiple of their applications, of which 148 (48\%) actually did obfuscate one or multiple applications. While the majority of developers (253, 82\%) had heard of reverse engineering, software plagiarism (201, 65\%) and software repacking (189, 61\%) and felt that Android applications in general were severely threatened by plagiarism and malicious repacking, they had the impression that their own applications were less likely to face those threats than apps in \textquote{general} (cf.\ Figure~\ref{fig:study:likert_combined}).
}

\item[Reasons to obfuscate.]{%
The following results are reported for 101 developers who voluntarily specified reasons for using obfuscation in a free text answer. 63 developers (62.3\%) used obfuscation to protect their intellectual property against malicious reverse engineering and theft. Interestingly, 14 (13.9\%) participants used {ProGuard} only because it came pre-installed with Android Studio and was easy to use. 18 (17.8\%) participants needed {ProGuard}'s optimization features and stated that adding obfuscation was trivial.
4 (4\%) participants apparently mis-understood the concept of obfuscation and enabled {ProGuard} to provide their users some extra level of security similar to encrypting files or using secure network connections. 7 (6.93\%) configured obfuscation because there was a policy (either given by the company they worked for or a customer) that dictated its use.}

\item[Verifying that obfuscation works:]{%
The following results are reported for the 69 participants who gave a free text answer on their method of verifying the success of obfuscating their app. 48 (69.6\%) developers verified the correct use of obfuscation by decompiling the application and manually looking for obfuscation features (e.g. obfuscated package, class or methodnames).
Six (8.7\%) participants relied on the Android Studio toolchain and interpreted no warning or error messages as successful obfuscation. Four (5.8\%) participants checked their apps' logfiles to verify their obfuscation. Finally, six (8.7\%) other participants verified obfuscation by comparing the size of the non-obfuscated with the obfuscated version of an application.}

\item[Reasons to not obfuscate.]{%
Out of the 185 developers who gave reasons to not obfuscate in a free text answers, 81 (54.8\%) thought about obfuscation and then decided against using it because they saw no reason to protect their application(s) against malicious reverse engineering, either because they open sourced their applications (17) or included no valuable intellectual property (64). 52 (35\%) participants tried to use obfuscation and gave up because they felt overwhelmed by ProGuard's complexity. They could not get third party libraries working or had other issues such as non-working JavaScript interfaces. Five (3.2\%) tried to understand the concept of obfuscation but failed. Eight (5.8\%) participants mentioned company policies that did not allow them to obfuscate code. However, no one elaborated on those policies in more detail.}

\item[Use of Obfuscation Tools.]{%
Furthermore, 148 participants gave details on the obfuscation tools they had used. Most of them (127; 85.8\%) had used {ProGuard}. 12 participants (8.1\%) used the Jack toolchain\footnote{The Jack toolchain was deprecated in March 2017 (cf.~\url{https://android-developers.googleblog.com/2017/03/future-of-java-8-language-feature.html})} 
11 participants (7.4\%) used {DexGuard} and 6 participants (4\%) used {ReDex}. 4 participants mentioned other less popular obfuscation tools with only one appearance, like an obfuscation tool built into the Unity engine. Overall, 144 (97.3\%) of the participants had used {ProGuard} or similar tools.}
\end{description}
        
\subsection{Discussion}
The survey results indicate a widespread awareness of the existence of obfuscation tools among Android developers, the consideration of using obfuscation and the actual use of obfuscation. ProGuard emerged as the most prominent tool to obfuscate. 
We also learned that many Android developers suffer from misconceptions (e.g. using obfuscation to secure network connections) and seem to be overwhelmed by using obfuscation correctly (e.g. the inability to obfuscate an app, but exclude certain components from obfuscation).
Generally, we also observed the lack of a threat model: one participant explicitly stated \textquote{I wasn't sure my apps would be even popular enough so that someone would bother to copy them. If they would get popular, I'd release an update with obfuscation on.} Many developers did not see a reason to obfuscate their own app(s) despite being aware of an abstract risk. One participant explicitly spoke of their experiences with piracy, stating \textquote{I see it as highly unlikely, that someone is actually interested in reverse engineering my code. However, I have encountered several fraud cases as an Android developer. All consisted of minimum reverse engineering efforts, i.e. people decompiled my app, changed the advertising ID code, repacked it, and published it under a different name.} 
We find that the lack of concrete threat models explains a low motivation to obfuscate; to obtain a better understanding of the barriers to obfuscation, we decided to investigate the usability issues mentioned by a substantial number of participants in depth.

\section{Obfuscation Experiment}
\label{sec:study}
The large scale measurement study and developer survey described above raised an interesting paradox: Roughly half of our survey participants claimed to have tried obfuscation in the past, but only 25\% of the apps in our measurement study were obfuscated.
We hypothesized that this discrepancy may be explained by the fact that developers may \emph{attempt} obfuscation, but be unsuccessful due to difficulties in using their obfuscation tool.

To test this hypothesis that the leading obfuscation tool might suffer from \emph{usability problems}, we conducted an online experiment to investigate how developers interact with the {ProGuard} obfuscation framework.

\paragraph{Study Design}
We designed an online, within-subjects study to compare how effectively developers could quickly write correct, secure {ProGuard} configurations. Again, we recruited developers with demonstrated Android experience from Google Play.
Participants were assigned to complete a short set of Android obfuscation tasks, using {ProGuard}. All participants completed the same set of two {ProGuard} tasks. After finishing the tasks, participants completed a brief exit survey about the experience. We examined participants' submitted {ProGuard} configuration for functional correctness and security. The study was approved by our institutions' ethics review boards (see Appendix~\ref{app:ethics} for more details).

\emph{Why {ProGuard}:}
We chose to use {ProGuard} as the obfuscation tool for our experiment for two reasons: first, it comes pre-installed with Android Studio, the standard IDE for Android application development. Second, our online survey participants overwhelmingly used ProGuard.

\paragraph{Recruitment and Framing}
Similarly to our survey, we recruited Android developers from Google Play to participate in our developers study. We emailed \studyinvites developers in batches, asking them to volunteer for a study exploring how Android developers use {ProGuard} to obfuscate apps. We did not mention security or privacy in the recruitment message. We assigned each invitee a unique pseudonymous identifier (ID) to allow us to link their study participation to Google Play metadata without being able to de-identify them.
Recipients who clicked the link to participate in the study were directed to a landing page containing a consent form. After affirming they were over 18, consented to the study, and were comfortable with participating in the study in English, they were introduced to the study, given access to an Android Study project containing our skeleton app and instructions (including screenshots) on how to import it and set it up.
We also provided brief instructions for using the study infrastructure, which we describe next.

\paragraph{Experimental Setup}
After reading the study introduction, participants were instructed to work on the tasks themselves. Our aim was to have developers write and test {ProGuard} configurations. We wanted to capture the {ProGuard} configuration and the Android application code that they typed. To achieve this, we prepared a Gradle based Android application development project for Android Studio as a skeleton, compressed the project to a zip file and provided a download link. We asked participants to download the zip file, import the project into their Android Studio development environment, work on the tasks, put their solutions in a new zip file and upload this file to our study server. After uploading the solution's zip file, we provided a link to the exit survey that allowed us to connect the {ProGuard} solutions to the survey responses.

\subsection{The Tasks}
\label{subsec:study:second}
To investigate possible usability issues with {ProGuard}, we aked participants to use {ProGuard} to complete two obfuscation tasks on the skeleton app we provided in the zip file.

We designed tasks that were short enough so that the uncompensated participants would be likely to complete them before losing interest, but sufficiently complex to offer insights into the usability of {ProGuard}. Most importantly, we designed tasks to model real world problems that Android developers using {ProGuard} could reasonably be expected to encounter in their professional career.
We chose both tasks after investigating {ProGuard} centered {StackOverflow} discussions and {GitHub} repositories. Both tasks are amongst the most popular {ProGuard} related discussions on {StackOverflow} and represent the most popular modifications in {ProGuard} configuration files on {GitHub}.

For each task, participants were provided with stub code and some commented instructions. These stubs were designed to make the task clear and ensure the results could be easily evaluated, without providing too much scaffolding. We also provided Android application and {ProGuard} code pre-filled so participants could test their solutions.

\paragraph{Task 1 - Configure:}
The first task required participants to activate {ProGuard} within the default Gradle configuration file. The goal was to fully obfuscate the Android application.

Participants were asked to solve this task so we could investigate their ability to complete a basic {ProGuard} configuration. Possible errors include the inability to activate obfuscation at all or a misconfiguration of {ProGuard} that disables obfuscation.

\paragraph{Task 2 - Obfuscate and Keep:}
\begin{sloppypar}
The second task required developers to configure {ProGuard} to obfuscate one specific class (\emph{SecretClass}) of our skeleton app, while keeping a second class (\emph{OpenClass}) and its function (\emph{doStuff()}) unobfuscated. To solve this task, developers were expected to use {ProGuard}'s ``\texttt{-keep}'' flag for the \emph{OpenClass} class. 

The challenge for this task was to correctly use the ``\texttt{-keep}'' flag. Depending on the specified arguments, developers could potentially leave the \emph{SecretClass} unobfuscated or obfuscate \emph{OpenClass} instead.
\end{sloppypar}

\paragraph{Exit Survey}
\label{subsec:study:exitsurvey}
Once both tasks had been completed and the zip file was uploaded, participants were directed to a short exit survey.\footnote{We used LimeSurvey for this; the full questionnaire is available in the Appendix.} We asked for opinions about the tasks they had completed, their assessment of their configurations for both tasks and general questions related to obfuscation and reverse engineering and their previous experience with {ProGuard} and other Android obfuscation tools.

\subsection{Evaluating Solutions}
\begin{sloppypar}
We used the code submitted by our participants for each task, henceforth called a solution, as the basis for our analysis.
We evaluated the correctness of each participant's solution to each task. Every solution was independently reviewed by two coders, using a codebook prepared ahead of time based on the official {ProGuard} configuration documentation. Differences between the two coders were adjudicated by a third coder.

We assigned correctness scores to valid solutions only. To determine a correctness score, we considered several different {ProGuard} parameters. A participant's solution was marked correct (1) only if their solution was acceptable for every parameter; an error in any parameter or a parameter that weakened the {ProGuard} configuration security resulted in a correctness score of 0.

To assess the correctness of Task 1, we evaluated the {Gradle} and {ProGuard} flags in participants' solutions.
Whenever participants enabled {ProGuard} using both the ``\texttt{minifyEnabled true}'' and ``\texttt{proguardFiles proguard-rules.pro}'' options in the configuration file, we rated the solution correct.
Solutions that did not specify one of these options or included the ``\texttt{-dontobfuscate}'' flag were rated incorrect.

For Task 2 correctness, we evaluated whether participants enabled obfuscation for the \emph{SecretClass} class and its \emph{doSecretStuff()} method but left the \emph{OpenClass} class and its method \emph{doStuff()} unobfuscated.
Similar to Task 1, we required participants to enable obfuscation by using the ``\texttt{minifyEnaled true}'' and the ``\texttt{proguardFiles proguard-rules.pro}'' options.
Additionally, correct solutions had to specify one of the following options ``\texttt{-keep}'', ``\texttt{-keepclassmemebers}'', ``\texttt{-keepclasseswithmembers}'', ``\texttt{-keepnames}'', ``\texttt{-keepclassmembernames}'', or ``\texttt{-keepclasseswithmembernames}'' for both the \emph{OpenClass} class and the \emph{doStuff()} method without including the \emph{SecretClass} and its \emph{doSecretStuff()} method.
Solutions that did not meet these criteria were considered incorrect.
\end{sloppypar}

\subsection{Results}
\begin{sloppypar}
In total, we sent \surveyinvites{} email invitations. Of these, 999 (1.9\%) requested to be removed from our list, a request we honored.

766 people clicked on the link in the email. Of these, a total of 280 people agreed to our consent form; 202 (72.1\%) dropped out without taking any action. We received zip files from the remaining 78 participants. We excluded eight submissions from further evaluation: one participant submitted a broken zip file, five submitted zip files without a {ProGuard} configuration file included, two submitted unmodified {ProGuard} configuration files.

The remaining 70 participants proceeded through at least one {ProGuard} task; of these, 66 started the exit survey, and 63 completed it with valid responses. Unless otherwise noted, we report results for the remaining 63 participants, who proceeded through all tasks and completed the exit survey with valid responses. Almost all (60, 95\%) of our participants had heard of the concept of software obfuscation before, and 54 (85\%) had been using {ProGuard} at least for one Android application in the past. 

Most participants (49, 77\%) mentioned an abstract threat of reverse engineering or malicious repackaging for Android applications in general; however, similarly to the online survey we conducted in Section~\ref{sec:survey} only a small number of participants estimated a high risk for malicious repackaging for their own app(s).

Surprisingly, all of the 70 participants who changed the configuration for Task 1 submitted a correct solution by adding both the ``\texttt{minifyEnabled true}'' and ``\texttt{proguardFiles proguard-rules.pro}'' options.

Task 2 was correctly solved by only 17 (22\%) participants, all of which could solve both tasks in a correct way. Of the 53 incorrect solutions for Task 2, 30 solutions did not include the \emph{-keep} option for the \emph{OpenClass} class. These mistakes resulted in obfuscated classes that should be kept unobfuscated. 17 of the 53 incorrect solutions did include the \emph{-keep} option but misspelled the package name for the \emph{OpenClass} class. Six of the 53 incorrect solutions included the wildcard option for class names which disabled obfuscation for the \emph{SecretClass} class.

41 of our participants rated their own solutions as correct. However, only 11 of them actually submitted correct solutions for both tasks. Overall, 52 participants self-reported previous experience with {ProGuard} of which 13 correctly solved both obfuscation tasks. Only one of the 11 participants with no previous {ProGuard} experience was successful.
\end{sloppypar}

\paragraph{Discussion}
We found that all participants, regardless of their experience with {ProGuard}, were able to solve the trivial task to obfuscate the complete app with {ProGuard}.
However, we found a low success rate for the task that required more complex configuration, which substantiated the usability problems mentioned in our developer survey. Being unfamiliar with {ProGuard} use essentially disqualified participants from being able to configure partial obfuscation. Critically, participants were unable to verify whether {ProGuard} had been configured correctly; i.e. whether obfuscation had been successful. These results underline a critical usability problem with {ProGuard} that likely contributes to the lack of obfuscation in the wild. 

\section{Discussion}
\label{sec:discussion}
To our knowledge, this paper is one of the first comprehensive analyses of software obfuscation in the Android ecosystem.
While earlier work relating to software obfuscation in Android apps focused on reversing the effects or the detection of certain structures despite obfuscation, our work investigates the prevalence of obfuscation in general and the awareness among developers for potential threats and benefits.

\paragraph{Security through insignificance?}
Our large-scale analysis showed that the majority of developers do not take the basic steps to protect their apps.
Even for the most popular apps with upwards of 10,000,000 downloads, high risk candidates for obfuscation-related threats,
the intentional obfuscation percentage remains below 50\%.
In our studies, participants assigned a low threat-potential for obfuscation-related attacks to their apps while assuming a greater threat-level for the whole app ecosystem.
Through provided write-ins we learned that many developers perceive their apps as too insignificant to ever fall prey to intellectual property theft or plagiarism.
This ``security through insignificance''-approach could prove fatal to the ever increasing number of small developers in the Android ecosystem.

\paragraph{Optional obfuscation:}
Another factor that seemingly contributes to the unwillingness of developers to use provided obfuscation tools
is the complexity for certain tasks.
The unwillingness is based on a low base motivation to begin with,
stemming from the negligible perceived personal threat,
in combination with cryptic error messages and confusing documentations as soon as tasks increase in complexity.

A certain mind-set seems to have contributed further to the  rejection of obfuscation:
some participants voiced concerns that obfuscation would destroy their  ``completed'' applications.
This view of obfuscation usage as an optional -- not essential --  development practice could play a larger role in hampering the acceptance of software obfuscation among developers.

\paragraph{Recommendations:}
Our findings indicate that there are two critical problems preventing
widespread adoption of obfuscation in the Android ecosystem.
The first is technical, and may have a technical solution: {ProGuard} is
difficult to use correctly. We believe that it may be possible to
automatically detect complicating factors (like {WebView} use) and automatically
generate valid {ProGuard} configurations for developers. If successful, this
would allow obfuscation to be enabled by default within Android Studio and
other development environments. The second problem is that developers are not
motivated to deploy obfuscation given a low perceived risk and high perceived
effort. Developers also view obfuscation as an optional, possibly
``app destroying'' step instead of an integral part of the build process.
While improved interfaces and automation for obfuscation may improve the
perceptions of effort, more research and education regarding the risks of
plagiarism is needed.
A technical solution may take the form of new obfuscation techniques or obfuscations applied by the market instead of relying on developers to protect themselves, their users, and the ecosystem at large.

\section{Threats to Validity}
\label{sec:limitations}
In this section, we detail issues that may have affected the validity of our results and the steps we have taken to ensure that our results are as accurate as possible.

\paragraph{App Analysis}
Our dataset of 1.7 million apps was downloaded from public accessible Google Play Android apps.
This is a common methodology, and like all similar studies we run the risk that paid apps or apps in other markets have different properties. These populations (paid apps in particular) may have additional incentives to obfuscate. However, we believe that the high overlap of apps that are available as both free and paid apps, and identical apps available in multiple markets, minimizes this risk.

Our choice of measuring main package obfuscation is not perfect;
it is possible that a developer does not obfuscate the main package but obfuscates the remainder of the app.
To estimate the frequency of this practice,
we examine how many apps without main package obfuscation have obfuscated packages that do not have multiple occurrences in the overall dataset.
We found that only 22,868 apps (1.30\% of all apps in the dataset) meet this criteria.
This establishes an upper bound on the error of this heuristic.
We note that an alternative approach to main package analysis would  have  been  to remove  third-party  library  packages  after  identification with obfuscation-resistant library detection tools such as \textsc{LibRadar}~\cite{ma2016}, \textsc{LibScout}~\cite{backes2016}, or \textsc{LibD}~\cite{li2017}.
This whitelist approach to package filtering would by design miss new or rarely used libraries, so we opted for the conservative approach of main package analysis.

\paragraph{Online Survey and Developer Study}
As with any user study, our results should be interpreted in context. We chose an online study because it is difficult to recruit ``real'' Android application developers (rather than students) for an in-person lab study at a reasonable cost. Choosing to conduct an online study resulted in less control over the study environment, but it allowed us to recruit a geographically diverse sample.

Because we targeted developers, we could not easily take advantage of services like Amazon's Mechanical Turk or survey sampling firms. Managing online study payments outside such infrastructures is very challenging; as a result, we did not offer compensation and instead asked participants to generously donate their time. As might be expected, the combination of unsolicited recruitment emails and no compensation may have led to a strong self-selection effect, and we expect that our results represent developers who are interested and motivated enough to participate. However, as the recruitment in Section~\ref{sec:survey}
demonstrates, while our participants have higher average app ratings, they overall cover a representative sample of Google Play developers, both in app popularity and frequency of updates. 

In any online study, some participants may not provide full effort, or may answer haphazardly. In this case, the lack of compensation reduces the motivation to answer in a manner that is not constructive; those who are not motivated will typically not participate. We attempted to remove any obviously low-quality data (e.g., responses that are entirely invective) before analysis, but cannot discriminate perfectly.

\section{Related Work}
\label{sec:relwork}
Software obfuscation has been studied  as defense against reverse engineering~\cite{collberg1997},
to prevent intellectual property attacks~\cite{collberg2002},
as disguise for malware~\cite{you2010},
and to avoid user profiling~\cite{ullah2014}.
Researchers successfully employed code obfuscation techniques to avoid detection tools,
including anti-malware software~\cite{protsenko2013,zheng2013,rastogi2014},
repackaging detection algorithms~\cite{huang2013},
and app analysis tools~\cite{hoffmann2016},
although performance of anti-malware software improved in a more recent study~\cite{maiorca2015}.
A number of works detail different obfuscation techniques in general~\cite{collberg1997,borello2008,collberg2009,you2010},
 for the Java programming language~\cite{chan2004,sakabe2005,hou2006}, and for Android apps in particular~\cite{ghosh2013,protsenko2013,faruki2014}.
Other work relating to obfuscation in Android apps has focused on  reversing the effects of obfuscation~\cite{udupa2005,bichsel2016} or on detecting certain features of an app despite obfuscation, like code reuse~\cite{hanna2013,zhang2014,linares-vasquez2014,glanz2017}, the detection of repacked malware~\cite{linn2003,faruki2013,spreitzenbarth2013,garcia2015}, or identification of third-party libraries~\cite{ma2016,backes2016}.

Previous Android developer studies were performed in the context of privacy, Trusted Layer Security/Secure Sockets Layer (TLS/SSL) security, and cryptographic Application Programming Interfaces (APIs).
Balebako et al.\ performed interviews and online surveys to investigate how app developers make decisions about privacy and security, identifying several hurdles and suggesting improvements that would help user-privacy~\cite{balebako2014a,balebako2014b}.
Jain et al.\ suggested design changes to the Android Location API based on the results of a developer lab study~\cite{jain2014}. 
Fahl et al.\ and Oltrogge et al.\ conducted developer surveys and interviews, revealing deficits in the handling of TLS/SSL and suggesting several improvements~\cite{fahl2012,fahl2013,oltrogge2015}.
Nadi et al.\ found in a study that Java developers struggle with perceived low-level cryptography APIs~\cite{nadi2016}. 
Concerning obfuscation on the Android platform,
Ceccato et al.\ assessed in experiments the impact of Java code obfuscation on the code comprehension of students, finding that obfuscation delays, but not prevents tampering~\cite{ceccato2009,ceccato2014}.
Pang et al.\ surveyed 121 developers about their knowledge concerning app energy consumption~\cite{pang2016}. 
Compared to these works,
our root cause analysis focuses on obfuscation knowledge and ability to use the obfuscation tool {ProGuard} among Google Play developers.
Related to a previous developer study investigating the impact of information sources on code security by Acar et al.~\cite{acar2016},
we find that developers are generally knowledgeable about the benefits and basic configuration,
but fail to correctly perform the process for more complex setups.

Finally, in a pre-print concurrent with our work, Dong et al.\ also investigate the use of obfuscation in the Android ecosystem~\cite{dong2018}. While that work is solely focused on technical measurements of obfuscation (similar in focus to our Sections~3 and 4), our research works with the developers responsible for obfuscation to determine the root causes of why apps are or are not obfuscated. Our app measurements are  more comprehensive (\allanalyzedapps{} apps from Google Play market vs.\ 114,560 apps) and use measurement techniques grounded in specifications of the most common obfuscation tools (instead of machine learning approaches). 

\section{Conclusion}
\label{sec:conclusion}
This paper presents the first comprehensive evaluation of the state of software obfuscation for benign Android applications. We built \toolname{} to analyze the use of obfuscation in \allanalyzedapps{} free Android applications available in Google Play. Our investigation reveals that \obfuscatedapps{} were obfuscated by their developers, leaving more than 75\% unprotected against malicious repacking. In an online survey with \surveyvalid{} Google Play developers, 78\% of the participants had heard of obfuscation while only 48\% actually used software obfuscation -- more than 85\% of the participants used {ProGuard} -- in the past. Interestingly, the majority of the participants recognized that software obfuscation in general is a laudable approach to protect against malicious repackaging. However, only few of them saw a reason to protect their own apps. Finally, in a within-subjects study with \studyvalid{} real Android developers, we learned that 78\% of the participants could not correctly complete a realistic {ProGuard} obfuscation task. Participants who self-reported no previous experience with {ProGuard} had a negligible chance to correctly obfuscate the study application beyond the trivial option to obfuscate it entirely.

Overall, our studies show that the current use of software obfuscation for benign Android applications leaves manifold challenges for future research. We find that both misconceptions about software obfuscation many of our participants suffered from and the challenges in using {ProGuard} correctly seem to be the root cause for the low adoption rate of software obfuscation in the Android ecosystem. Hence, future research needs to find more effective ways to make the concept and relevance of software obfuscation concepts accessible to Android developers and has to work on a more efficient and usable integration of software obfuscation tools.

\bibliographystyle{acm}
\bibliography{refs.bib,related-work.bib}

\appendix\label{sec:appendix}
\section{Ethical Considerations}
\label{app:ethics}
% Included in the appendix

We conducted two user studies in the context of this paper. Both the survey presented in section~\ref{sec:survey} and the developer study in section~\ref{sec:study} were approved by the ethical review board of University A in Germany and by the Institutional Review Board of University B in the US. Additionally, the strict data and privacy protection laws in Germany were taken into account for collecting, processing and storing participants' data.
Our userstudies were targeted towards Android developers who had made their app public by offering it on Google Play. For ecological validity reasons we decided against recruiting local computer science students. To reach this rather specific group of Android developers, we gathered email addresses from developers who had published apps on Google Play from their public Google Play profiles. We selected a random sample and emailed them an invitation to one of our studies (This participant recruitment procedure is in line with work by Acar et al.~\cite{acar2016}). Our invitation email included a link to our website, where they could access information about the purpose of our research, a consent form that explained how participant data would be used and a contact form. The email further included a link to be blacklisted; hashes of the blacklisted email addresses are shared across several research groups participating in similar developer studies.

\section{Online Survey}
\label{appendix:questionnaire}
\begin{description}
\item[General Questions]{\hfill
\begin{itemize}
\item Which of these have you heard of in the context of Android apps? Please check all that apply.\\(Reverse Engineering, Repackaging of Software, Software Plagiarism, Obfuscation)
\item How likely do you think Android apps are \ldots \\(Reverse Engineered, Repackaged, Software Plagiarism, Obfuscated), scale: (Very Unlikely, Unlikely, Neutral, Likely, Very Likely, I don't know)
\item How likely do you think your *own* Android apps are... \\\emph{(Reverse Engineered, Repackaged, Software Plagiarism, Obfuscated), scale: (Very Unlikely, Unlikely, Neutral, Likely, Very Likely, I don't know)}

\item How much do you feel the intellectual property of your *own* Android apps is threatened by... \\\emph{(Reverse Engineering, Software Plagiarism), scale: (Very Unlikely, Unlikely, Neutral, Likely, Very Likely, I don't know)}
\end{itemize}
}

\item[Terminology]{\hfill
\begin{itemize}
\item Reverse engineering is:\\ \emph{(Translate binary files to source code, Translate source code to binary files, Analysis of pure source code, Analysis of binary files,  Reconstruction of app logic, Testing an app's functionality, I don't know, Other [with free text])}

\item Reverse engineering can be used for:\\ \emph{(Understanding an app's logic, Circumvention of licence or security checks, Repackaging of an app, Stealing IP addresses, Attacks on Android users who have your app installed, Remote attacks on mobile phones, I don't know, Other [with free text])}

\item Software plagiarism is:\\ \emph{(Repackaging existing software and rebranding it as your own, Use of third party open source code in your software, Imitating software to trick users, Copy pasting code found on the internet, I don't know, Other [with free text])}

\item Software plagiarism can be used for:\\ \emph{(Obtaining software revenue, Distributing disguised malware, Attacking users that have your app installed, Attacking distribution services, I don't know, Other [with free text])}

\item Obfuscation is:\\ \emph{(Making source code unreadable or difficult to understand so only authorized developers can work on it, Making source code unreadable or difficult to understand before compilation, Hiding binaries from the user, Preventing acces to the deployed application, I don't know, Other [with free text])}

\item Obfuscation can be used for:\\ \emph{(Making reverse engineering more difficult, Prevent others from attacking vulnerabilities within your application, Hiding the logic within your application, Optimization of app performance, I don't know, Other [with free text])}

\item Have you heard of obfuscation before?\\ \emph{(Yes, No, Uncertain)}
\item Have you ever thought about using obfuscation?\\ \emph{(Yes, No, Uncertain)}
\item Did you obfuscate at least once before?\\ \emph{(Yes, No, Uncertain)}
\end{itemize}
}

\item[Obfuscation tools]{\hfill
\begin{itemize}
\item Please select all Android obfuscation tools that you have heard of prior to this study.\\ \emph{(ProGuard, DexGuard, Jack, DashO, ReDex, Harvester, Other [with free text])}

\item Please select all Android obfuscation tools that you have used before.\\ \emph{(ProGuard, DexGuard, Jack, DashO, ReDex, Harvester, Other [with free text])}

\item Please select all Android obfuscation tools that you have actively decided against using.\\ 
\emph{(ProGuard, DexGuard, Jack, DashO, ReDex, Harvester, Other [with freetext])}

\item Which tools do you use to remove unused library code?\\
\emph{(ProGuard "Minify", Android Studio "Minify", I remove it manually, I never remove unused library code from my apps, Other [with free text])}
\end{itemize}
}

\item[Obfuscation 1]{\hfill
\begin{itemize}
\item How did you first encounter obfuscation? \\\emph{[Free text]}

\item How many apps have you worked on? \\\emph{[Number input]}
\item How many of those where obfuscated? \\\emph{[Number input]}
\item Why did you use obfuscation on those apps? \\\emph{[Free text]}
\item Why did you decide against obfuscating apps? \\\emph{[Free text]}
\item Did you verify that obfuscation was successful? \\\emph{(yes, no)}
\item How did you verify if obfuscation was successful? \\\emph{[Free text]}
\item Why did you decide against using obfuscation? \\\emph{[Free text]}
\end{itemize}

}
\end{description}

\subsection{{ProGuard} Study - Exit Survey}
\label{appendix:exit_survey}

After completing the programming task, developers were asked to fill out a final survey.

\begin{description}
\item[Tasks]{\hfill
\item Do you think you solved the tasks correctly? \\\emph{(Task1, Task2), scale: (Yes, No, I don't know)}

\item Do you have additional comments on the tasks? \\\emph{[Free text]}
}

Followed by the \texttt{General Questions, Terminology and Obfuscation tools} question groups from the online survey (cf. Appendix~\ref{appendix:questionnaire})

\item[ProGuard]{\hfill
\begin{itemize}
\item What do you use Proguard for? \\\emph{(Testing, Minifying Code, Optimization, Obfuscation)}

\item After using Proguard, how did you verify that it achieved its goal? \\\emph{(I do not verify that Proguard worked, Reverse Engineering, Other [with free text])}

\item Why have you never used Proguard before? \\\emph{(No need, Never heard of it, Too complicated, I have other tools, Other [with free text])}
\end{itemize}
}
\end{description}

\end{document}